# Methane Emissions from Super-emitting Coal Mines in Australia quantified using TROPOMI Satellite Observations

*Pankaj Sadavarte*\*,†, *Sudhanshu Pandey*†, *Joannes D. Maasakkers*†, *Alba Lorente*†, *Tobias Borsdorff*†, *Hugo Denier van der Gon*‡, *Sander Houweling*†,§, *Ilse Aben*†

†SRON Netherlands Institute for Space Research, 3584 CA, Utrecht, The Netherlands.

‡Department of Climate, Air and Sustainability, TNO, 3584 CB, Utrecht, The Netherlands.

§Department of Earth Sciences, Vrije Universiteit, Amsterdam, 1081 HV, Amsterdam, The Netherlands.

**Corresponding Author\***

E-mail: p.sadavarte@sron.nl

**Abstract**

Two years of satellite observations were used to quantify methane emission from coal mines in Queensland, the largest coal producing state in Australia. The six analyzed surface and underground coal mines are estimated to emit 570±98 Gg a-1 in 2018-2019. Together, they account for 7% of the national coal production, while emitting 55±10% of the reported methane emission from coal mining in Australia. Our results indicate that for two of the three locations our satellite-based estimates are significantly higher than reported to the Australian government. Most remarkably, 40% of the quantified emission came from a single surface mine (Hail Creek) located in a methane-rich coal basin. Our findings call for increased monitoring and investment in methane recovery technologies for both surface and underground mines.

**Introduction**

Methane ($CH_4$) is the second most important greenhouse gas and is responsible for 25% of the anthropogenic radiative forcing in the atmosphere.[1] Due to its shorter atmospheric life time (~12 years) compared to $CO_2$ and higher greenhouse warming potential, the mitigation of methane emissions is an efficient method to tackle the near-term climate warming.[2] The current methane growth rate, however, challenges existing climate policies, including the Paris Agreement (PA), and will ask for additional reductions on top of what is already foreseen to attain the PA goals.[3] To do this in an efficient manner, an improved understanding and quantification of anthropogenic methane emissions is of vital importance.

The fossil fuel industry, including oil/gas (O/G) production and coal mining, accounts for one-third of the total anthropogenic methane emission.[4, 5] Coal mining is responsible for about 12% of total anthropogenic methane emissions[4, 5], with 90% coming from underground mines.[6] The recent Global Methane Budget suggests an increase of 38% (12 Tg) in emissions from coal mines between 2000-2009 and 2017[4, 7], most likely due to the increase in global coal production. Methane emission from coal mines have been quantified using atmospheric measurements from ground-based and aircraft campaigns.[8, 9] Space-borne remote sensing instruments have been used to detect and quantify methane emissions on a regional scale and can provide a measurement-based integral quantification of large point sources.[10-13] Recent developments in space-borne instruments with sub-kilometer pixel resolution have made identification and quantification of emissions from individual oil and gas facilities and coal mine shafts possible.[14, 15] However, these high-resolution satellites have limited spatial coverage as they tend to only observe targeted areas.[15]

Here we use satellite observations of the TROPOspheric Monitoring Instrument (TROPOMI) onboard the Copernicus Sentinel-5 Precursor (S-5P) satellite, launched on 13 October 2017. It is a push-broom imaging spectrometer in a sun-synchronous orbit providing daily global methane columns ($XCH_4$) with a local overpass time at 13:30.[16] The daily global coverage combined with fine spatial resolution of 7 × 7 km$^2$ (7 × 5.5 km$^2$ since August 2019) of TROPOMI enables the detection of super-emitters of methane in a single overpass.[12, 14, 17]

In this study, we quantify fugitive methane plumes from coal mines observed with TROPOMI over Queensland state in Australia (Figure 1). We use two years (2018-2019) of clear-sky column averaged methane (XCH$_4$) observations with the data driven cross-sectional flux method (CSF) to estimate emissions. This method has been used in previous studies to quantify emissions from point sources using satellite observations.[12, 14, 15] We compare our estimates with coal mine emissions from a global inventory and those officially reported by Australia to the United Nations Framework Convention on Climate Change (UNFCCC).[18] The study highlights the super-emitter behavior of three coal mines or coal mine clusters. The identification and quantification of integrated overall methane fluxes from coal production sites using satellite observations can help to further improve the national inventory and prioritize emission reduction targets.

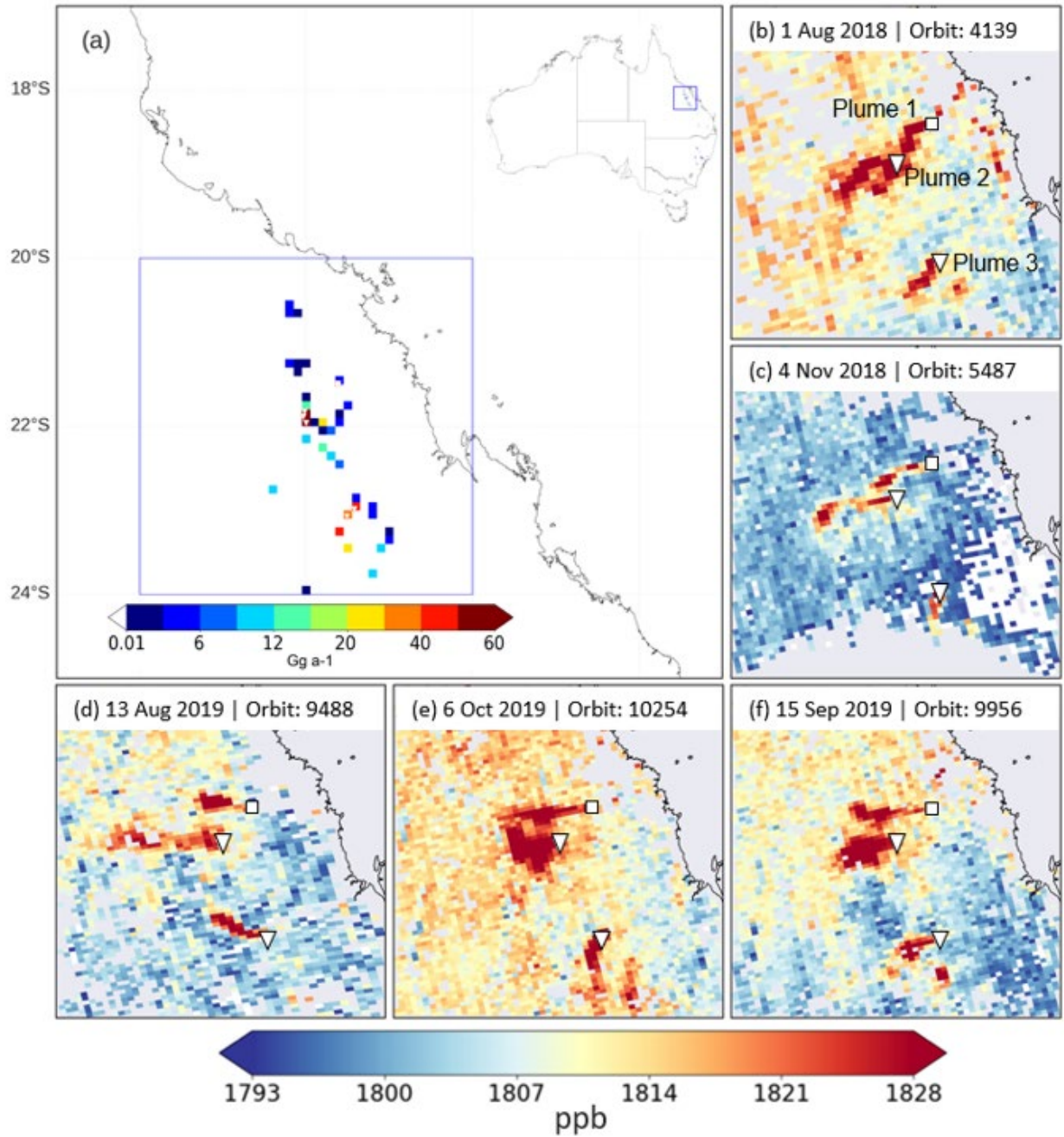

**Figure. 1: TROPOMI observations and methane emissions over the study domain.**
Panel (a) shows a 0.1° × 0.1° gridded map of reconstructed bottom-up methane emissions from coal mines in Queensland, Australia (*19*). The blue square ranging from latitude 20°–24°S and longitude 146°–150°E indicates the domain containing the three source locations of our study. The inset panel shows the map of Australia and relative location of the study domain which lies in the north-east. Examples of the persistent XCH$_4$ plumes observed are shown for different TROPOMI orbits over the study domain (b-f) during 2018 and 2019. The surface mine at source 1 is identified by the square at the origin of the top plume and the underground mines at source 2 and 3 are indicated with triangles near the middle and the bottom plumes. Cloud free observations are mostly found during the months of June till November in both years. TROPOMI methane column (XCH$_4$) is given in ppb and the gridded methane emissions inside the study domain are given in Gg a$^{-1}$.

**Materials and Methods**

**TROPOMI observations.** The TROPOMI scientific data product used here was retrieved using the RemoTeC full-physics algorithm with improvements that resulted in a more stable retrieval and correction for surface albedo biases.[20] Total column methane (XCH$_4$) is retrieved with nearly uniform sensitivity in the troposphere from its absorption band around 2.3 μm and 0.7 μm using earthshine radiance measurements from the shortwave infrared (SWIR) and near infrared (NIR) channel of TROPOMI.[20-22] This new dataset has shown good agreement with the measurements from the well-established Total Carbon Column Observing Network (TCCON)[23] and with the Greenhouse gases Observing SATellite – GOSAT.[24] The TROPOMI XCH$_4$ measurements used in this analysis were screened for cloud-free coverage and low aerosol content using the quality flag provided in the data products (we use qa = 1). Data quality qa = 1 signifies XCH$_4$ is filtered for solar zenith angle (<70°), viewing zenith angle (<60°), smooth topography (1 standard deviation surface elevation variability < 80 m within a 5 km radius) and low aerosol load (aerosol optical thickness < 0.3 in the NIR band). The TROPOMI data was corrected for XCH$_4$ variations due to surface elevation by adding 7 ppb per km surface elevation with respect to the mean sea level.[25] TROPOMI XCH$_4$ data show artificial stripes in the flight direction most probably due to swath position dependent calibration inaccuracies, which were corrected by applying fixed mask de-striping approach to the L2 data developed for the TROPOMI XCO retrieval.[26, 27]

For emission quantification from TROPOMI detected plumes, orbits from 2018 and 2019 were screened with > 500 individual observation pixels in the domain of 20°–24°S and 146°–150°E (Figure 1a). To ensure that emission quantifications are not influenced by systematic surface albedo or aerosol bias we reject orbits that show high correlation (|R| > 0.5) of XCH$_4$ with surface albedo or aerosol optical thickness. Seventy-five orbits containing a total of 124 clear-sky observations over the three sources were thus selected and used for emission quantification. The temporal spread shows most observations in the months of July-December in both 2018 and 2019 (Figure 2). The presence of clouds during January till June limits the availability of TROPOMI during these months. However, quarterly raw coal production numbers in 2018 and 2019 show variations of less than 5%, so we expect only minor differences in emission rates over the year.

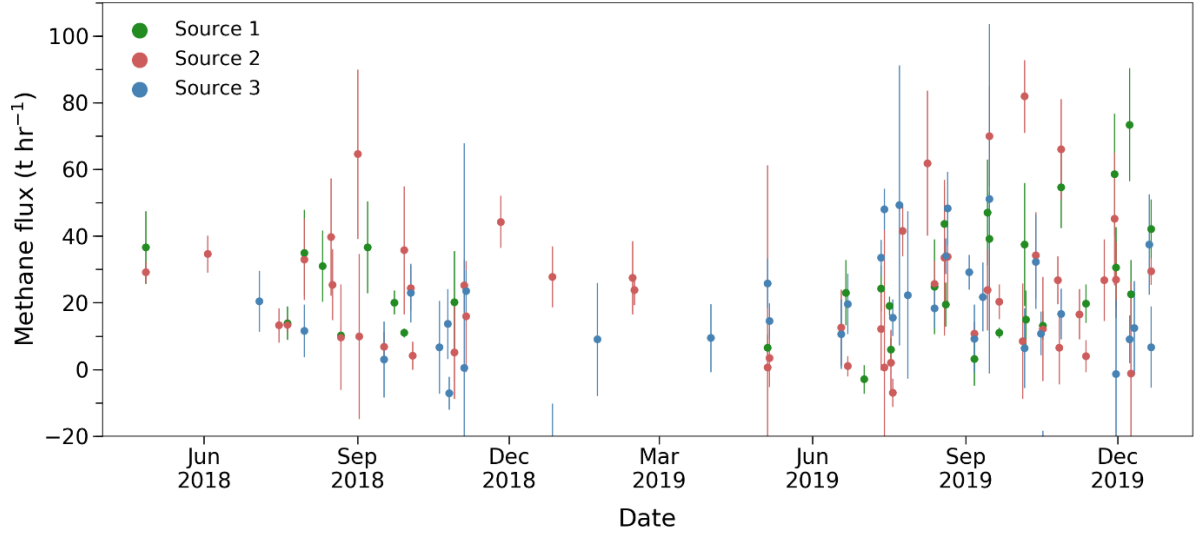

**Figure 2. Methane emissions fluxes quantified from individual TROPOMI observations.**
Daily methane flux estimates derived from TROPOMI observations for the three sources that
were used for the annual quantification. A total of 124 clear-sky scenes spanning over the source
areas from 75 orbits are shown here. The methane source rate for each XCH$_4$ plume is given with
its uncertainty (1σ).

**Cross-sectional flux method.** We quantify methane emissions from TROPOMI observations
using the cross-sectional flux method[28] as shown below in equation (1).

$$Q = \bar{C}U_{eff} \text{ where, } \bar{C} = \frac{1}{n} \times \sum_{j=1}^{n} \int \Delta\Omega(x_j, y)dy \quad (1)$$

where, the source rate $Q$ (t hr$^{-1}$) is calculated as the product of the integrated methane column
enhancement $\bar{C}$ and the effective wind speed $U_{eff}$. The methane column enhancement $\Delta\Omega$ $(x_j, y)$ is
computed by sampling the plume using transects orthogonal to the plume direction (y-axis) in the
downwind of the source (x-axis) (Figure S1). The sampled observations are integrated across each
transect within the limits defined by the length of the transect. For a daily source rate, we take the
mean of all the emission estimates calculated for individual transects (*j=1…n, where n is number
of transects*) between the source and the end of the plume. For deriving the effective wind speed
($U_{eff}$), we use the pressure weighted average boundary layer wind speed $U_{blh}$ from ERA5
meteorology. Varon et al.[14] derived a relationship between $U_{eff}$ and $U_{blh}$ for TROPOMI
observations as $U_{eff} = (1.05 \pm 0.17)$ $U_{blh}$ by using the Weather Research and Forecasting model
coupled with chemistry (WRF-Chem) where modelled methane emissions were compared with
the cross-sectional flux estimates. For our case, we have assumed $U_{eff}$ equal to $U_{blh}$.

Transects across the plume have been defined for each source by estimating the downwind direction and dimensions of the plume. We start with a smaller rectangular mask of dimension (*length × breadth*) 0.4° × 0.2° placed at the source in the downwind direction inferred from boundary layer average ERA5 meteorology, to define the area containing the plume (Figure S1). Next, we rotate this mask from -40° to +40° at 5° intervals around the inferred ERA5 wind direction such that the average XCH$_4$ enhancement in the rectangular mask is maximal. After we set the new wind direction, the length of the rectangular mask in the downwind direction (along x-axis) is varied to define the end of the plume. This end is fixed by incrementing the length of rectangular mask by 0.1° intervals until the difference between methane enhancement of two consecutive increments is less than 5 ppb. Similarly, the width of the rectangular mask (along y-axis) was fixed by incrementing the width in lateral direction of the plume at an interval of 0.05° until the incremental change in methane enhancement is less than 5 ppb.

We define 15 equally spaced transects between the source and the end of the rectangular mask for calculating the source rates. We ignore the first three transects due to their close proximity to the source, where XCH$_4$ may be underestimated due to partial pixel enhancement.[12, 14] To avoid underestimation of emissions due to incomplete sampling of the plume by a transect due to missing pixels, we only consider transects that have more than 75% overlap with TROPOMI pixels. With this requirement, we only calculate the source rate from plumes with at least three or more transects. The methane enhancements for each pixel along the transects is defined relative to the background XCH$_4$ which is calculated as the average of 0.5° × 0.5° area centered at a distance of 0.1° upwind from the source. If the number of background observations is less than 20, we use the median XCH$_4$ of all pixels in the domain (20°–24°S, 146°–150°E) as background XCH$_4$. To account for other emissions in the downwind plume, we subtract the contributions from surrounding coal mines[18, 19] (Figure S2b), the other anthropogenic sources from EDGARv4.3.2 global emissions[5] (Figure S3b) and emissions from oil and gas[29] (Figure S3c) within the plume for each source. In some cases, we estimate small negative emissions in Figure 2 possibly due to high XCH$_4$ values in the background. As the location of the background and source regions are shifting around the source with changes in daily wind directions, we expect this error to average out in the mean source rate. We compute the uncertainty in the daily emission rate by accounting for uncertainty in the mean enhancement, the pressure-weighted average boundary layer ERA5 wind speed and the uncertainty derived from $U_{eff}$ and $U_{blh}$ equation. (see Supporting Information, Section S1).

**Bottom-up emission estimates.** The bottom-up emissions from the global inventory of EDGARv4.3.2[5] (most recent year 2012) and the Australian national inventory reporting[30] (for 2018) were used in this study to compare with the TROPOMI emission estimates. EDGARv4.3.2 uses tier-1 (global default emission factors) and some tier-2 (region specific) information to estimate national emissions from all anthropogenic sources. These emissions are available on a 0.1° × 0.1° grid, allowing comparison with the observations. For this purpose, the 2012 EDGAR emissions from coal mines were scaled to 2018 using the ratio in coal production from 2012 to 2018 of Queensland state (the derived 2018 emissions are referred as EDGARv4.3.2*). As the location of EDGAR emissions for coal mines do not exactly match the locations of the sources studied here, the emissions in the grid cell closest to the source locations were chosen as representing these coal mine locations (Figure S3a). The Australian national inventory report (NIR) utilizes more detailed tier-2 and tier-3 (facility specific) methodologies but is not available at a resolution beyond the state level. The national inventory provides methane emissions from coal for the categories of surface mines and underground mines at state level.[30] For the emissions associated with the coal mines of study, we use gridded emissions from Sadavarte et al.,.[19] These emissions were estimated using grouped emissions in the surface and underground category at state level from the national inventory and distributed these to the respective surface and underground coal mines within the state using coal production of individual mines as a distribution proxy along with the gas content profiles of the coal basins.[19] Section S2 of the supporting information provides the link to access the data used in the analyses.

**Results and discussion**

**TROPOMI localization of emission sources.** For the three distinct plumes that are consistently visible in the TROPOMI methane data over the Bowen Basin in Queensland state, we use the wind-rotation technique described by Maasakkers et al.[31] combined with the reconstructed high-resolution bottom-up inventory by Sadavarte et al.[19] (Figure S2) to determine which sources are responsible for the enhancements. The wind-rotation method (see Supporting Information, Section S3) traces the location of a source by averaging TROPOMI data after aligning the observations from individual days with the local wind vector (from GEOS-FP 10 m)[32]. The source location is then determined by comparing the resulting averaged rotated downwind 'plumes' for a full grid of rotation points. For the most northern plume seen in TROPOMI, we identify the emission source to be the Hail Creek surface mine. The middle plume originates from the underground mines of Broadmeadow, Moranbah North and Grosvenor, and for the most southern plume, the Grasstree and Oaky North underground mines are responsible (see Supporting Information, Section S3). Given the limited spatial resolution of the TROPOMI observations and the close vicinity of the coal mines at the second and third source locations, we could not further distinguish the contributions of the individual mines. Table 1 summarize the details about the geographical location, mining type and production. Supporting Figure S4 shows the satellite imagery of the source locations.

**Table 1:** Source location details and methane emission quantification using TROPOMI observations.

| Details | Source 1 | Source 2 | Source 3 |
|---|---|---|---|
| **Location** | Hail Creek | Broadmeadow, Moranbah North, and Grosvenor | Grasstree and Oaky North |
| **Mine type** | Surface | Underground | Underground |
| **Mining method** | Dragline, truck and shovel | Longwall | Longwall |
| **Total raw coal production in million tonnes** | 2018-19: 7.7<br>2019-20: 5.8 | 2018-19: 19.2[a]<br>2019-20: 19.0[a] | 2018-19: 13.7[b]<br>2019-20: 12.4[b] |
| **Longitude, Latitude** | 148.380°E, 21.490°S | 147.980°E, 21.825°S<br>147.967°E, 21.885°S<br>147.996°E, 21.962°S | 148.579°E, 22.988°S<br>148.486°E, 23.072°S |
| **Number of clear sky observations in TROPOMI** | 32 | 54 | 38 |
| **Annual emissions using CSF method (Gg a-1) [μ ± 2σ]** | 230 ± 50 | 190 ± 60 | 150 ± 63 |

[a]includes raw coal production from Broadmeadow, Moranbah North and Grosvenor underground coal mines.
[b]includes raw coal production from Grasstree and Oaky North underground coal mines.

**TROPOMI methane emission quantification and uncertainty estimate.** For the emission quantification, we screen individual TROPOMI orbits for sufficient spatial coverage over the region (20°–24°S and 146°–150°E), source locations, data-quality indicators, as well as favorable wind speed conditions. Figure 1 shows a few typical observations with signals from the three source locations clearly visible in the data. For each selected orbit, methane emissions are quantified for each source location using the cross-sectional flux method.[28] In this method, emissions are calculated by taking the product of line integrals of methane enhancements and wind speed, perpendicular to the downwind direction of the methane plume, similar to Varon et al.[14]. A total of 124 plumes from 75 screened orbits have been quantified for the period 2018-2019 (Figure 2). We use the average boundary-layer ERA5 wind speed for the TROPOMI overpass time of 04:00 UTC. Figure 2 shows the temporal variability in the methane flux from the three source locations with uncertainty of one standard deviation on each source rate. We estimate relative uncertainties of 55% on average (range of 18%-98%) on the daily emission source rates for non-negative enhancements. These uncertainties include the standard deviation in the different transects used in the CSF; the uncertainty in the background by varying the area it is calculated over; and the uncertainty in the wind speed by using windspeeds within $\pm$ 2 hours of the overpass time (see Supporting Information, Section S1). The largest uncertainties are caused by the presence of high methane in the background making it difficult to isolate the mine's signal and cases with low wind speeds as influences from turbulent transport become important which are not accounted for in our method.[28] Therefore, estimates at wind speeds below 2 m s$^{-1}$ are excluded. The number of days with emission quantifications is mainly limited by the presence of cloud cover but although there is quite some variation in the daily estimates and the error on each methane flux, the number of observations in combination with the random sampling over a 2-year period is representative of the methane source and sufficient to quantify annual emissions.[15]

The combined annual methane emission from the three persistent (more than 75% of the 124 screened orbits had high methane enhancements downwind of the source locations) sources is estimated at 570 $\pm$ 98 Gg a$^{-1}$ (Figure 3). Multiple sensitivity tests confirm the robustness of our emission estimate within its uncertainty (see Supporting Information, Section S1, Figure S5, Table S1). Together, the three sources emit a factor of 7 more than their bottom-up estimates in the global EDGARv4.3.2* emission inventory (84 Gg a$^{-1}$)[5]. Our estimate is also higher by a factor 2 compared to the reconstructed high-resolution bottom-up (RBU) emissions from the national inventory report, (250 Gg a$^{-1}$)[18, 19]. There is reasonable agreement between the national methane emission from coal mines reported by EDGARv4.3.2 (1228 Gg a$^{-1}$ for 2012) and the national inventory report for 2018 (972 Gg a$^{-1}$). The large difference in emissions between the three sources in these two inventories (Figure 3) is most likely explained by the different spatial proxies used for the disaggregation of national methane emissions (Figure S2b and S3a). The EDGARv4.3.2 global inventory[5] uses coal production activity from the World Coal Association and spatial proxies from the Global Energy Observatory for all countries other than the United States (USGS coal mines), Europe (EPRTRv4.2) and China[33]. While the Sadavarte et al.[19] – inventory uses Australian UNFCCC NIR reported emissions at the state level and spatially distributes these emissions using coal mine locations from the Queensland state web portal.[34] In short, EDGAR distributes the emissions over a much larger number of locations and it is not surprising that for the individual locations a discrepancy is found. Since the coal mine locations of the Queensland state web portal were also verified from the mining operation reports of coal mine companies, we believe these locations to be the most reliable.

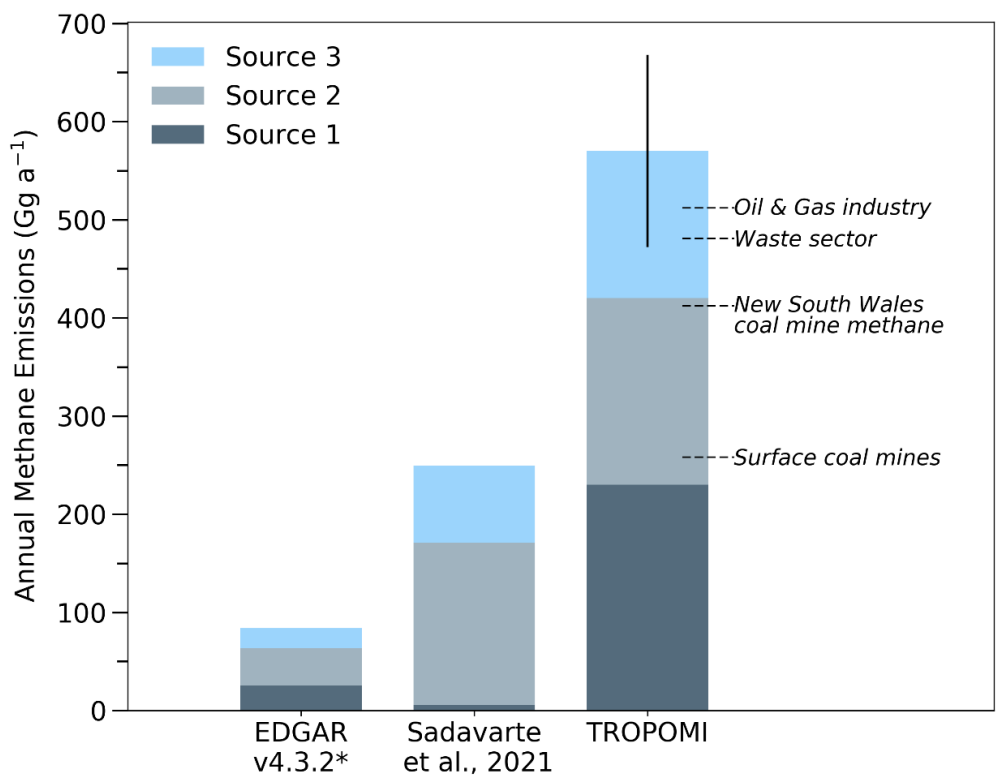

**Fig. 3: Annual methane emissions for three coal mine sources.**
Annual methane emissions estimates for the coal mine sources of the persistent plumes observed in TROPOMI data. The left bar shows the annual methane emissions from the global inventory of EDGARv4.3.2 available for 2012. EDGARv4.3.2* indicates the projected emissions for 2018 calculated after accounting for the change in coal production in Queensland state in 2018 relative to 2012. The middle bar shows the reconstructed bottom-up emissions from Sadavarte et al.[19] for the three sources using national emissions communicated to UNFCCC for 2018 and proxies such as coal production for individual mines and gas content profile. The right bar shows the total annual emissions estimated using TROPOMI observations for the period 2018-2019. The error bar represents 2σ uncertainty (95% confidence interval). Total emissions from TROPOMI are also compared with nationally reported greenhouse gas emissions from selected sectors and categories of Australia for 2018 using the dashed horizontal lines on the TROPOMI bar.

Focusing on the individual sources, our estimate for Hail Creek is more than 35 times the reconstructed bottom-up emission[19] (RBU: 6 Gg a$^{-1}$, TROPOMI: 230 ± 50 Gg a$^{-1}$) and 15% higher than the reported methane emission from all surface mines in Queensland state combined (196 Gg a$^{-1}$) (Table S2). Our Hail Creek estimate accounts for 88% of Australia's total reported surface coal mine emissions, suggesting a large underreporting of methane emissions in the national inventory reporting for surface mines (Figure 3, Table S3). Similarly, emissions from Grasstree and Oaky North underground mines are a factor 2 higher[19] (RBU – 79 Gg a$^{-1}$, TROPOMI – 150 ± 63 Gg a$^{-1}$), while emissions from the Broadmeadow, Moranbah North and Grosvenor mines are consistent with the reconstructed estimate[19] (RBU – 165 Gg a$^{-1}$, TROPOMI – 190 ± 60 Gg a$^{-1}$).

**Comparing emissions with national estimates.** Applying the cross-sectional flux method to two years of TROPOMI observations we estimate a total methane source strength of 570 ± 98 Gg a$^{-1}$ for the three source locations, equivalent to an average methane flux of 65 ± 11 t hr$^{-1}$. This can be broken down to 230 ± 50 Gg a$^{-1}$ $CH_4$ emissions from source 1 (a single surface mine) and 340 ± 86 Gg a$^{-1}$ $CH_4$ from source 2 and 3 (five underground mines). To put these emissions in national context, we compare them to Australian methane emissions from other source sectors. Our estimate for these three coal mine sources represent over 10% of the total reported methane emission from Australia in 2018, and exceeds the emission from the oil and gas industry sector (512 Gg a$^{-1}$) as well as the entire waste sector (480 Gg a$^{-1}$) (Figure 3, Table S3). The six mines produce only 7% of the national raw coal production (41 million tonne) but represent 55% of the national methane emissions from coal production reported for 2018 (Table S2 and S3). The Hail Creek mine alone emits 20% of the national $CH_4$ emission from coal mining, while accounting for only 1% of the national coal production.

**Analyzing the TROPOMI derived emission factor for Australian coal mines.** Australia, and in particular the state of Queensland, is known for its production of liquified natural gas (LNG) by extracting coal seam gas (CSG) from the methane-rich Bowen and Surat basins, which is also being exported internationally since 2015.[35] The gassy nature of the underground mines in Queensland state is well established, and led to infrastructure not only to release methane to the atmosphere through ventilation shafts, but also to capture and utilize it for power generation or flare or transfer off-site (see Supporting Information, Section S4). Australia reports methane emissions from underground mines using a tier-3 Intergovernmental Panel on Climate Change (IPCC) accounting method, using country-specific methodologies and respective mine-specific measured emissions factors (see Supporting Information, Section S5). These tier-3 emissions are not disclosed publicly for individual mines, but grouped and reported at state level in the national inventory report.[18, 30] (Queensland state produced 51% of the raw coal and emitted 56% of the national fugitive methane from coal mines.[18, 30] (Supporting Information, Table S2)). This hampers direct verification of mine-specific emissions using atmospheric measurements, like those from TROPOMI. In the case of surface mines, methane emissions are likely unabated and escape to the atmosphere throughout the mining operations. Although as per NGER guidelines "venting or flaring of in-situ gas can also occur from open cut coal mines", it is less common and less efficient since the coal seam is in direct contact with the atmosphere, providing a diffusion pathway that is difficult to capture. Moreover, combustion of large gas volumes with low $CH_4$ content is more expensive than with higher concentrations. For national inventory reporting, these emissions are calculated using a mix of tier-2/tier-3 emission factors and coal production data.[30] The tier-3 emission factors in Australia are measured following the National Greenhouse and Energy Report guidelines[36] for each surface mine in the Gunnedah, Western, Surat, Collie, Hunter and Newcastle basins only. The surface mines in the Bowen basin, including Hail Creek, uses a tier-2 basin-average emission factor (1.2 m$^3$ CH$_4$/tonne of raw coal) from William et al.[37].[30] It is difficult to assess how representative this tier-2 approach is for the local situation but our results indicate that it leads to a severe underestimation in the case of Hail Creek.

The emission factor inferred from TROPOMI data for the underground mines 2 and 3 amounts to 10-11.50 g CH$_4$ per kg raw coal, consistent with emission factors from EDGARv4.3.2, IPCC default values and Kholod et al.[6] for mining at 200-400 m depth (Table S4). Whereas the national and state level emission factors for underground mines (for 2017 and 2018) are 25%-50% lower than TROPOMI based implied emission factor (Table S4). Lower country-specific emission factors compared to IPCC defaults in itself are not surprising as local coal type and mitigation measures play an important role but we notice that especially for the mines of source 3 they are not in line with the TROPOMI based observations. For surface mine Hail Creek (source 1), the TROPOMI-inferred emission factor is 34 g CH$_4$ per kg raw coal, 22 times higher than the average of the IPCC default for < 200 m and Kholod et al.[6], i.e. 0.2, 0.52 and 2.03-3.38 g CH$_4$ per kg raw coal (Table S4).

**Understanding the super-emitting behavior of Hail Creek.** The Hail Creek mine was approved for an extension to highwall and underground mining activities in 2016.[38] Sentinel-2 satellite images over Hail Creek for 2018 to 2019 do not however show any significant change to the northeast of the surface mine, where the extension was proposed (see Supporting Information, Movie 1). The preparatory activities are seen to the northeast of the surface mine, suggest possible pre-mining degasification, starting before 2018. Typically, the degasification or pre-drainage is performed prior to underground mining as a safety measure against outbursts in the underground mine (see Supporting Information, Section S4). It involves draining the seam gas by either natural or active venting, combusting and/or flaring on-site or transferring off-site.[36] We do observe flaring activities over the extended area in July – September 2019[39] (Figure S6). However, no flaring activity was observed for the remainder of the analysis period in 2018-2019[39]. Most likely the TROPOMI-detected emissions at Hail Creek in 2018 and 2019 are due to surface mining and also possibly from pre-drainage activities.

In conclusion, to reduce the uncertainty in methane leakage from fossil fuel production, it is crucial to have accurate estimates of methane emissions from coal production. The TROPOMI instrument does not have the granularity of the ground-based measurements and/or monitoring of individual shafts as done by the mining companies. However, its observations provide a useful measure of emissions from the entire coal mine infrastructure, including emissions from ventilation shafts and other pre- and post- drainage systems like underground in-seam (UIS), surface to in-seam (SIS) and gas wells drilled for underground mines and any other unforeseen leakage. The good agreement for source 2 with the reconstructed bottom-up emissions shows that there can be good agreement with bottom-up reporting. When applying exactly the same method and approach to source 3 and source 1, however, we find large discrepancies with the reported values. The TROPOMI-inferred emission factor for source 3 (underground mines) is consistent with global studies and also with the value derived for source 2. On the other hand, for source 1 (surface mine Hail Creek) we find unexpected high emission for a surface coal mine and an implied emission factor which is more than an order of magnitude higher than any default factor in current IPCC guidelines for this source type. Overall, we find higher amounts of methane emitted, especially from the Hail Creek surface mine, pointing to the underreporting of Australian methane emissions to a level that would justify a revision of the national methane emission reported in the NIR to the UNFCCC. Our results show that satellite observations can provide a measurement-based integral quantification of an entire facility or production site. This is valuable complementary information next to emission estimates of individual processes or mine shafts. It can help to further improve national emission inventories and support identification of the most promising targets for mitigation.

## Author Contributions

P.S. and S.P. analyzed the TROPOMI data, performed the mass balance calculation and sensitivity studies with inputs from S.H. and I.A.; J.D.M. performed localization method for identification of coal mines; A.L. processed the operational data product of TROPOMI methane for 2018 and 2019; T.B. provided the support for de-stripping of TROPOMI orbits; P.S. and H.D.G. contributed to the bottom-up inventory analysis. P.S. wrote the manuscript with inputs from all the co-authors.

## Funding Sources

This work was supported through the GALES project (#15597) by the Dutch Technology Foundation STW, and the TROPOMI national program through NSO. P.S. and S.P. are funded through the GALES project (#15597) by the Dutch Technology Foundation STW, which is part of the Netherlands Organization for Scientific Research (NWO). A.L. and T.B. acknowledge funding from the TROPOMI national program through NSO.

## Acknowledgement

We thank the Earth Science Group team at SRON for developing the retrieval method for TROPOMI methane observation and consistent technical support throughout the period. We thank the team that realized the TROPOMI instrument and its data products, consisting of the partnership between Airbus Defense and Space Netherlands, KNMI, SRON, and TNO, commissioned by NSO and ESA. Sentinel-5 Precursor is part of the EU Copernicus program, and Copernicus Sentinel data of Scientific version for 2018–2019 have been used. We thank Prof. Bryce Kelly, University of New South Wales, for his continuous support and expert knowledge on coal mines in Australia. We acknowledge the provision of publicly available global bottom-up emission of greenhouse gases from EDGAR and meteorology data product of ERA5.

## Notes

The authors declare no competing interests.

# Methane Emissions from Super-emitting Coal Mines in Australia quantified using TROPOMI Satellite Observations

*Pankaj Sadavarte*[*,†]*, Sudhanshu Pandey*[†]*, Joannes D. Maasakkers*[†]*, Alba Lorente*[†]*, Tobias Borsdorff*[†]*, Hugo Denier van der Gon*[‡]*, Sander Houweling*[†,§]*, Ilse Aben*[†]

Supporting information includes:
      22 pages; 5 sections; 6 figures; 4 tables; and 1 animation

**Section S1: Uncertainty estimates and sensitivity analysis**

We compute the uncertainty in the daily emission rate by accounting for (a) the uncertainty in the mean enhancement due to variation in $XCH_4$ across each valid transect and in the upwind background (equation 1a), and (b) the uncertainty in the pressure-weighted average boundary layer ERA5 wind speed. We also account for an additional uncertainty of 16% (0.17/1.05) in the effective wind speed based on the relation $U_{eff} = (1.05 \pm 0.17)U_{blh}$ derived by Varon et al.,[14] These uncertainties are represented as relative standard deviations (RSD) i.e., the ratio of the standard deviation to the mean ($\sigma/\mu$) and added in quadrature to compute the uncertainty in the daily emission rate (equation 1b). To calculate the uncertainty in the two-year mean emission rates, we combine the uncertainty of individual emission rates as shown below in equation 1c:

$$\sigma_{enhancement} = \sqrt{\sigma_{transect}^2 + \sigma_{background}^2} \qquad (1a)$$

$$RSD_{ij} = \sqrt{RSD_{enhancement}^2 + RSD_{windspeed}^2 + RSD_{Ueff-Ublh}^2} \qquad (1b)$$

$$\sigma_j = \sqrt{\frac{\sum_i^n (Q_{ij} \times RSD_{ij})^2}{n^2}} \qquad (1c)$$

where $\sigma_j$ is the one standard deviation of the mean emission estimate of the $j^{th}$ source. $RSD_{ij}$ is the error on individual daily source rates $Q_{ij}$, where $i$ represents one out of $n$ orbits containing source $j$.

The uncertainty in the source rate includes uncertainties from the emission flux calculated across each valid transect in a plume as it reflects the variability in source enhancement sampled at different intervals. We estimate relative standard deviations as high as 98% in the daily emission source rate for non-negative enhancements. This uncertainty range was found to be even higher (>100%) for lower source rates, in this case less than 10 t hr$^{-1}$, similar to that found in Varon et al.[28]. Such high uncertainties can be explained by the selection of background observations, as other potential methane sources in the background area increase the spread of $XCH_4$ observations, increasing the uncertainty in the mean background $XCH_4$. Secondly, for <20 pixels in the upwind background, the median of the domain is used as mean $XCH_4$ background, which may introduce large uncertainties. The error in the wind speed has contributions from the hourly representation of the ERA5 winds, and the coarse resolution (0.25° × 0.25°) of the meteorology datasets. The relative standard deviation in daily source rates due to windspeed variations within ± 2 hours of the overpass time amounts to 48%.

To further test the robustness of our emission estimates, sensitivity tests have been performed varying the number of valid transects, calculating the enhancements by subtracting the median of the study domain instead of the upwind background mean, applying different quality flags and using the operational instead of the scientific TROPOMI XCH$_4$ data product. Supporting figure S5 shows that the total annual emission from the three sources are not sensitive to the number of valid transects used. As expected, we see a decrease in number of orbits considered for emission quantification as the required number of valid transects increases. Similarly, Table S1 tabulates the emissions computed using the operational data product of TROPOMI and relaxed quality flag qa > 0 combined with aerosol optical thickness < 0.1 and precision error < 10 ppb. We find that the annual emission estimates from these sensitivity tests to be in statistical agreement with our original emission estimate, i.e. the total emissions under varied conditions ranged from 500 – 565 Gg a$^{-1}$.

### Section S2: Data Availability

Different data sets used in the study are available from their respective data portals.

- TROPOMI methane Scientific product: https://ftp.sron.nl/open-access-data-2/TROPOMI/tropomi/ch4/14_14_Lorente_et_al_2020_AMTD/
- TROPOMI methane Operational product: https://s5phub.copernicus.eu/dhus/#/home
- ERA5 surface and pressure level hourly meteorology dataset: https://cds.climate.copernicus.eu/cdsapp#!/dataset/reanalysis-era5-single-levels?tab=form
- EDGARv4.3.2 global emission inventory of greenhouse gases : https://edgar.jrc.ec.europa.eu/overview.php/overview.php?v=432_GHG&SECURE=123

### Section S3: Details on plume rotation technique

The localization of the origin of the observed plumes is done by using the wind-rotation technique introduced by Maasakkers et al.[31]. TROPOMI observations on individual days are rotated around a suggested source location based on the local GEOS-FP 10 m wind vector[32] on that given day. If the suggested source location is indeed the main emission source, the downwind concentrations will always be larger than the upwind concentrations and an average of multiple days of rotated data will result in a mean 'plume' in the average. By repeating this exercise for a full grid of points, the resulting 'plumes' can be compared, the location with the largest enhancement compared to the background is estimated to be the actual emission source. Rotations for this study were performed using 2018-2019 TROPOMI methane data.[20]

For the most northern plume (source 1), the rotated data points towards 2km north and 2km east of the marked location (Figure S4) at the Hail Creek Surface mine, effectively pointing at the North-East corner of the mine. While the bottom-up emission estimates for this surface mine (RBU: 6 Gg a$^{-1}$) are rather low, no other sources are apparent in the vicinity ($\pm$ 3km) suggested by the rotation method as the footprint of the mine and pre-mining operations are much larger than that. The other two signals in TROPOMI appear to be caused by underground mines. For the source-2 plume, the rotation method leads us about 2km east of the Moranbah North mine (Figure S4) where multiple vents and drill-holes are visible in satellite imagery. Given the more spread out nature of the plumes observed here and the proximity of the Broadmeadow and Grosvenor mines, we are only able to analyze the aggregation of the three mines. The origin of the most southern source-3 is located 4km east of the Grasstree mine where most mining vents are visible, with a possible contribution from the Oaky North mine located to the south. Given the on-ground pixel resolution of the TROPOMI observations and the close vicinity of the coal mines for each of those two source locations, we could not further distinguish the contributions to individual mines.

In addition to the above identified mines, several surface mines are also found in the vicinity of the source 2 and 3 underground mines. The Goonyella Riverside (just above the Broadmeadow) and the Issac Plains ($\sim$13km to the east of Grosvenor) are one of the closest surface mines for source 2. Using the reconstructed bottom-up inventory[19], Goonyella Riverside contributes 13 Gg a$^{-1}$ and Issac Plains 2 Gg a$^{-1}$ $CH_4$ emissions annually. While Middlemount (16km to north of Grasstree), Lake Lindsay (16km to southeast of Grasstree) are the closest operating surface mines for source 3. They emit 3 Gg a$^{-1}$ $CH_4$ each annually in 2018. German Creek east and Oak Park surface mines are much closer to Grasstree, 10km to the east, but no active coal production is reported for them and hence we assume little or no emissions of relevance to the estimates here. Given the low estimates for the $CH_4$ emissions for these nearby surface mines, we can safely assume they do not contribute significantly to our TROPOMI emission estimates for sources 2 and 3.

### Section S4: Emissions reporting from underground and surface coal mines in Australia

Underground mines in Australia are equipped with ventilation and gas drainage systems more or less for the same reason. The ventilation fans and auxiliary shafts are installed to provide sufficient air circulation so as to dilute the contaminant air within the safety limits. While gas drainage systems are installed to reduce burden on the ventilation shafts carrying higher amounts of methane released during underground mining operations and to avoid outburst for personnel safety[40]. Effectively there are two types of gas drainage system, pre-drainage and post-drainage systems. The pre-drainage system involves extracting coal seam gas from the mine even before the mining operations have commenced. As quoted in the National Greenhouse and Energy Reporting for emission guidelines, "*Pre-gas drainage occurs prior to mining in an area, primarily to avoid safety hazards from potential outbursts that could result during extraction. Where the gas concentration in the coal seam is high, the gas needs to be drained prior to mining the strata.*".[36] While the post-drainage is performed during the mining operation that extracts not only methane but other gases from seam and surrounding strata[36]. Collectively this is also referred to as coal mine waste gas. The gas drainage systems for underground mines at source 2, Broadmeadow, Moranbah North[41] and Grosvenor and source 3, Grasstree[41] and Oaky North[42] include:

1) UIS – Underground in-seam

2) SIS – Surface to in-seam

3) Gas wells

In addition, under Carbon Credits (Carbon Farming Initiative – Coal Mine Waste Gas) Act 2011, the methane component of coal mine waste gas drawn from ventilation air and gas drainage system of underground mine is converted to carbon dioxide either by a flaring device or flameless oxidation device or electricity production devices[43]. The combustion of coal mine waste gas reduces the impact on global warming by converting it to carbon dioxide instead of direct release of methane emissions in the atmosphere (http://www.cleanenergyregulator.gov.au/ERF/Pages/Choosing%20a%20project%20type/Opportunities%20for%20industry/Mining,%20oil%20and%20gas/Coal-mine-waste-gas.aspx).      Given below is the list of power station operated by underground mines at source 2 and 3 that uses coal mine waste gas for generating electricity.

- Moranbah North – 63.9 MW operated by EDL since 2008. (https://edlenergy.com/project/moranbah-north/)

- Grosvenor, two power station – 21 and 15 MW operated by EDL since 2016. (https://edlenergy.com/project/grosvenor/)

- Oaky North – 15 MW operated by EDL since 2016. (https://edlenergy.com/project/oaky-creek-2/)

- Grasstree – 45 MW operated by EDL since 2006. (https://edlenergy.com/project/german-creek/)

As Australia follows the highest tier-3 UNFCCC methodologies for reporting methane from underground mines, the national emissions are further classified from ventilation air (649 Gg-emitted), post-mining handling (42.5 Gg-emitted), electricity generation (1.8 Gg-emitted; 261.5 Gg-captured) and flaring (3.9 Gg-emitted; 297.9 Gg-captured) activities of underground mines[30].

Similarly, for fugitive emissions from surface mines, the National Greenhouse and Energy Reporting (NGER) guidelines states, "*Venting or flaring of in-situ gas can also occur from open cut coal mines. This will be from surface in seam drainage (SIS). Such gas drainage from open cut* *coal mines is less common than for underground coal mines. The likelihood of significant in-situ gas in place prior to coal extraction is lower where targeted coal seams are closer to the surface."* It is clear that venting (to capture methane) or flaring of fugitive methane from coal seam can be a possible methane recovery technology for surface mines, be less efficient, because of rapid diffusion to the overhead atmosphere that it is in direct contact with.

## Section S5: Australia and its greenhouse gas reporting commitments

Australia is an Annex-I party to the UNFCCC, Kyoto Protocol and the Paris Agreement, under which it is obliged to report its greenhouse gas emissions each year, reduce its greenhouse gas emissions and track progress towards those commitments. Australia annually submits the National Inventory Reports to the UNFCCC. These emission estimates are compliant with UNFCCC reporting guidelines, IPCC 2006 guidelines for national greenhouse gas inventories, and supplementary reporting requirements under the Kyoto Protocol. (https://www.industry.gov.au/policies-and-initiatives/australias-climate-change-strategies/tracking-and-reporting-greenhouse-gas-emissions). Under the National Greenhouse and Energy Reporting (NGER) Scheme, entities such as facility or corporates that meet certain emission thresholds are obliged to report their emissions, energy production and energy consumption each financial year to the Clean Energy Regulator (http://www.cleanenergyregulator.gov.au/NGER/National%20greenhouse%20and%20energy%20reporting%20data/What-data-is-published-and-why). The emission threshold for a facility to report its emissions under the National Greenhouse and Energy Reporting Act 2007 is 25 ktonne or more of greenhouse gases ($CO_2$-e), production of 100 TJ or more of energy or consumption of 100 TJ or more of energy. Similarly, the threshold for corporate group is 50 ktonne or more of greenhouse gases ($CO_2$-e), production of 200 TJ or more of energy or consumption of 200 TJ or more of energy (http://www.cleanenergyregulator.gov.au/NGER/Reporting-cycle/Assess-your-obligations/Reporting-thresholds). The underground coal mining companies estimate methane emissions from 'mine return ventilation' and 'gas drainage to surface' activities.[36] For calculating the emissions, parameters like gas flow rate, proportion of methane in the gas stream, pressure of the gas (in case of gas drainage) at STP (standard temperature and pressure) and temperature of the gas at STP is measured. Using the molecular mass of methane, these are converted into the units of mass of emissions. While tier-3 measured emissions from surface mines follow methodology developed by Saghafi[44] which is also adopted by National Greenhouse and Energy Reporting (NGER). The primary data required for estimating emissions from surface mine includes parameters like in-situ gas content and the gas composition of coal. These are collected by drilling holes at multiple locations at a surface mine or a coal basin. These industry-reported emissions are compiled by the National Greenhouse Accounts department and reported as part of the national emissions to the United Nations Framework Convention on Climate Change (UNFCCC).

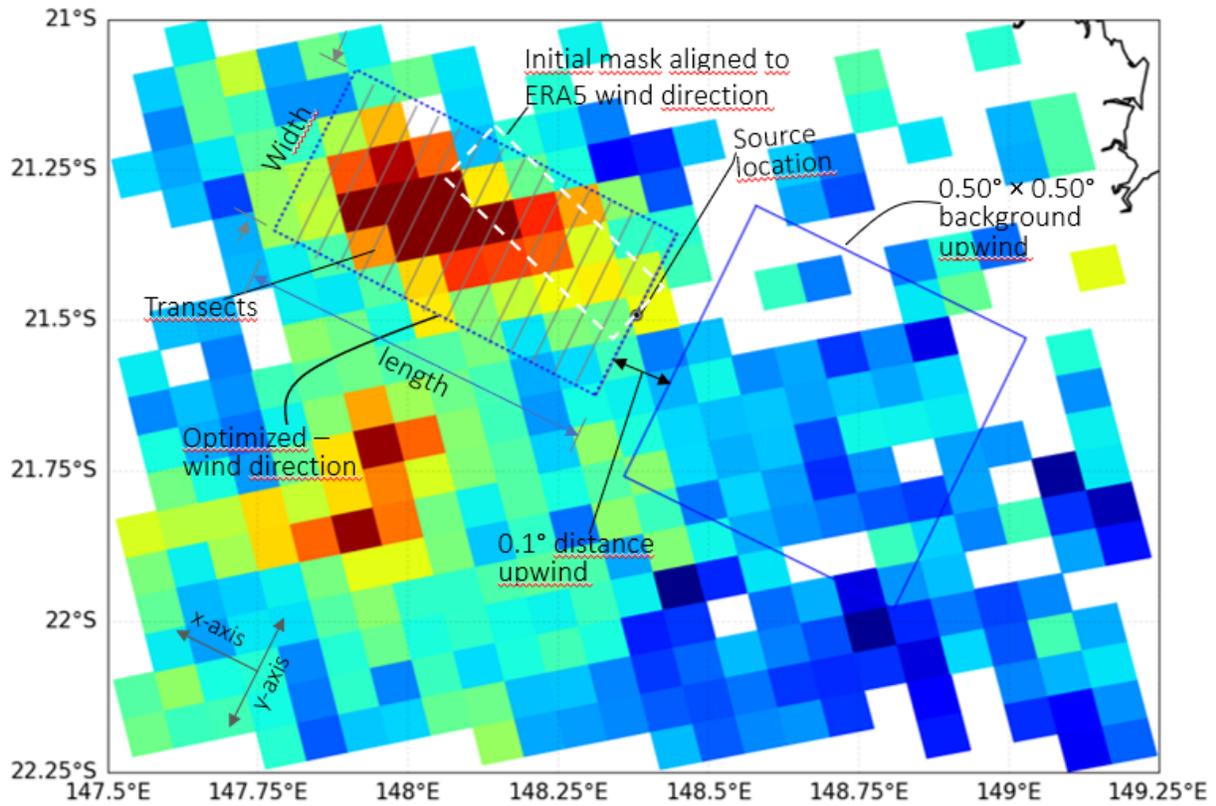

**Figure S1. Quantification method.**

Cross-sectional flux method showing the source location (black dot), upwind background area of 0.5° × 0.5° dimension (blue square), initial mask and wind direction (white dotted box), optimized wind direction over the plume (black dotted box), and the transects (grey lines).

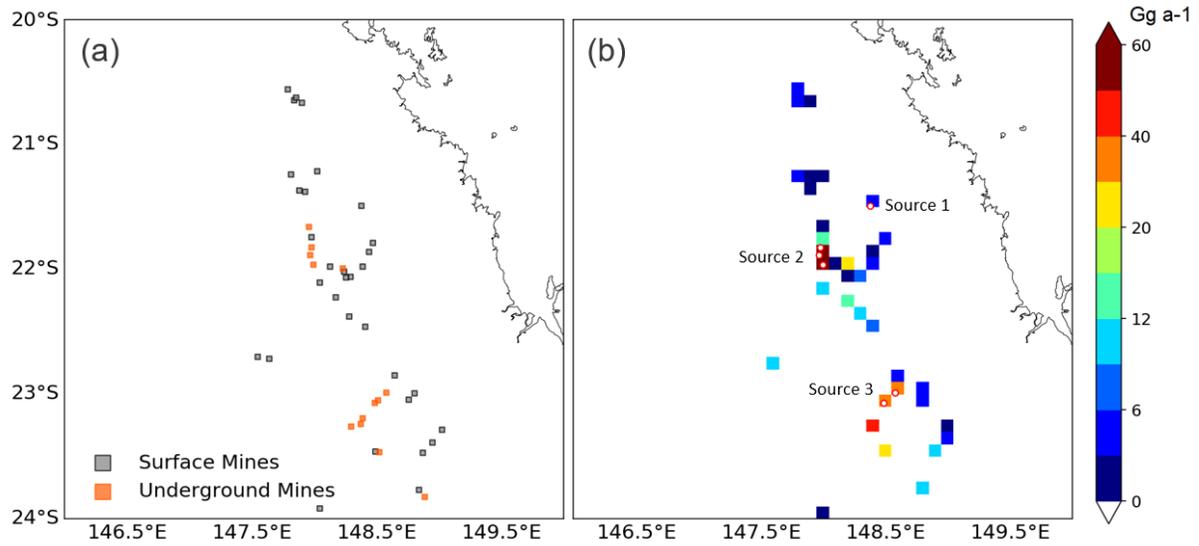

**Figure S2. Coal mines in Queensland state of Australia.**

Locations of surface and underground coal mines in Queensland, Australia and their methane emissions. (a) Active surface (grey) and underground (orange) coal mine locations (b) respective methane emissions gridded on 0.1° × 0.1° resolution from Sadavarte et al.,[19]. Red circles denote the location of the three source locations, source 1 – Hail Creek, source 2 – Broadmeadow, Moranbah North and Grosvenor and source 3 – Grasstree and Oaky Creek.

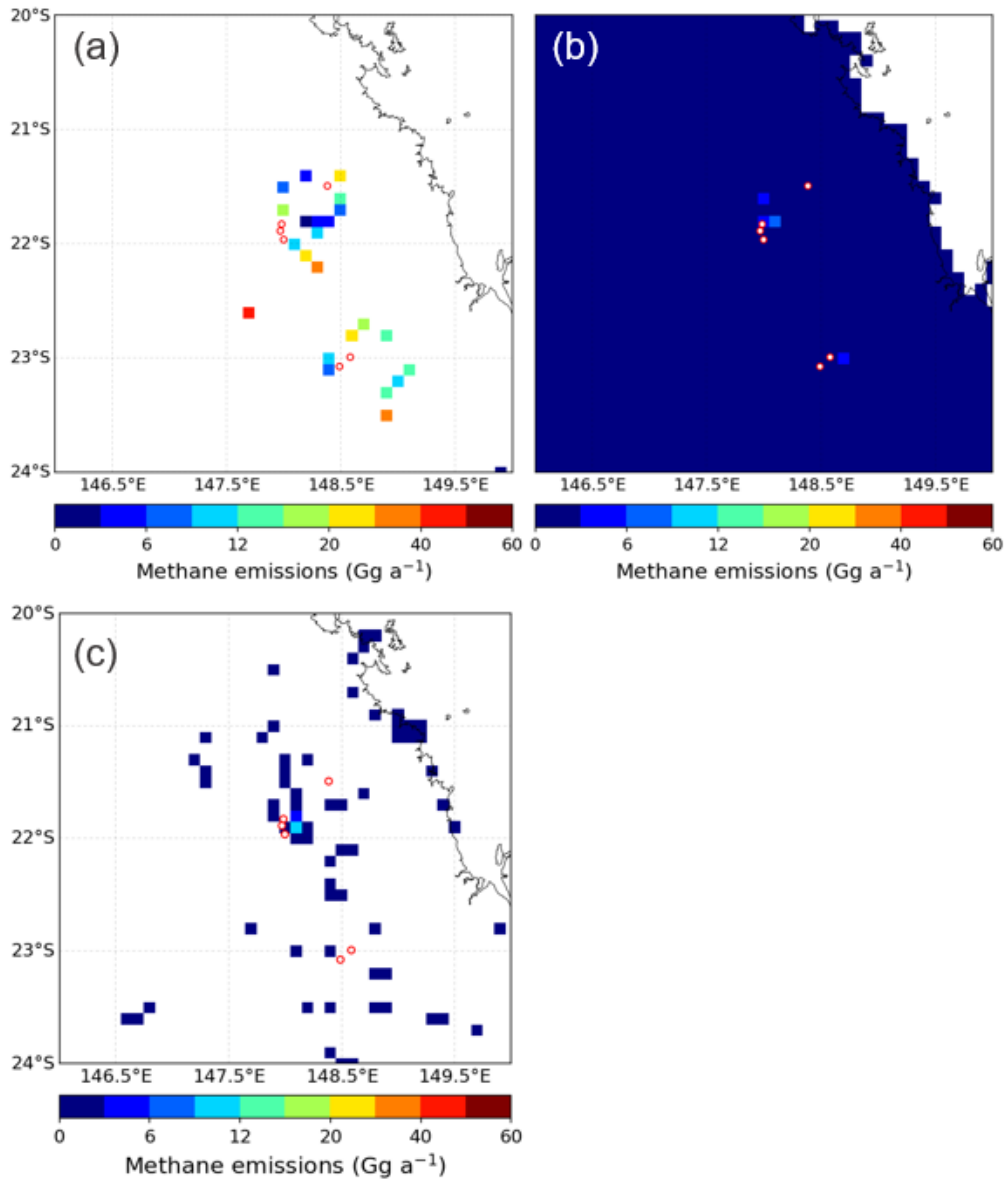

**Figure S3. Methane from bottom-up emission inventories over Queensland state.**

Methane emissions from the global inventory of EDGARv4.3.2 on a 0.1° × 0.1° grid resolution for a) coal mining activities, and (b) other anthropogenic sources of energy, transport, waste water, landfills, agriculture including livestock, paddy cultivation and others except coal mines and oil and gas category and (c) Oil and gas based methane emissions estimated from production, refining, processing, transport, and storage activities over the study domain in Queensland state on a 0.1° × 0.1° grid resolution from a global inventory by Scarpelli et al.,[29]. The EDGARv4.3.2 methane emissions are available for year 2012, while the oil and gas methane emissions from Scarpelli et al.,[29] is available for 2016. Red circles denote the locations of the three sources, Hail Creek (top), Broadmeadow, Moranbah North and Grosvenor (centre) and Grasstree and Oaky Creek (bottom).

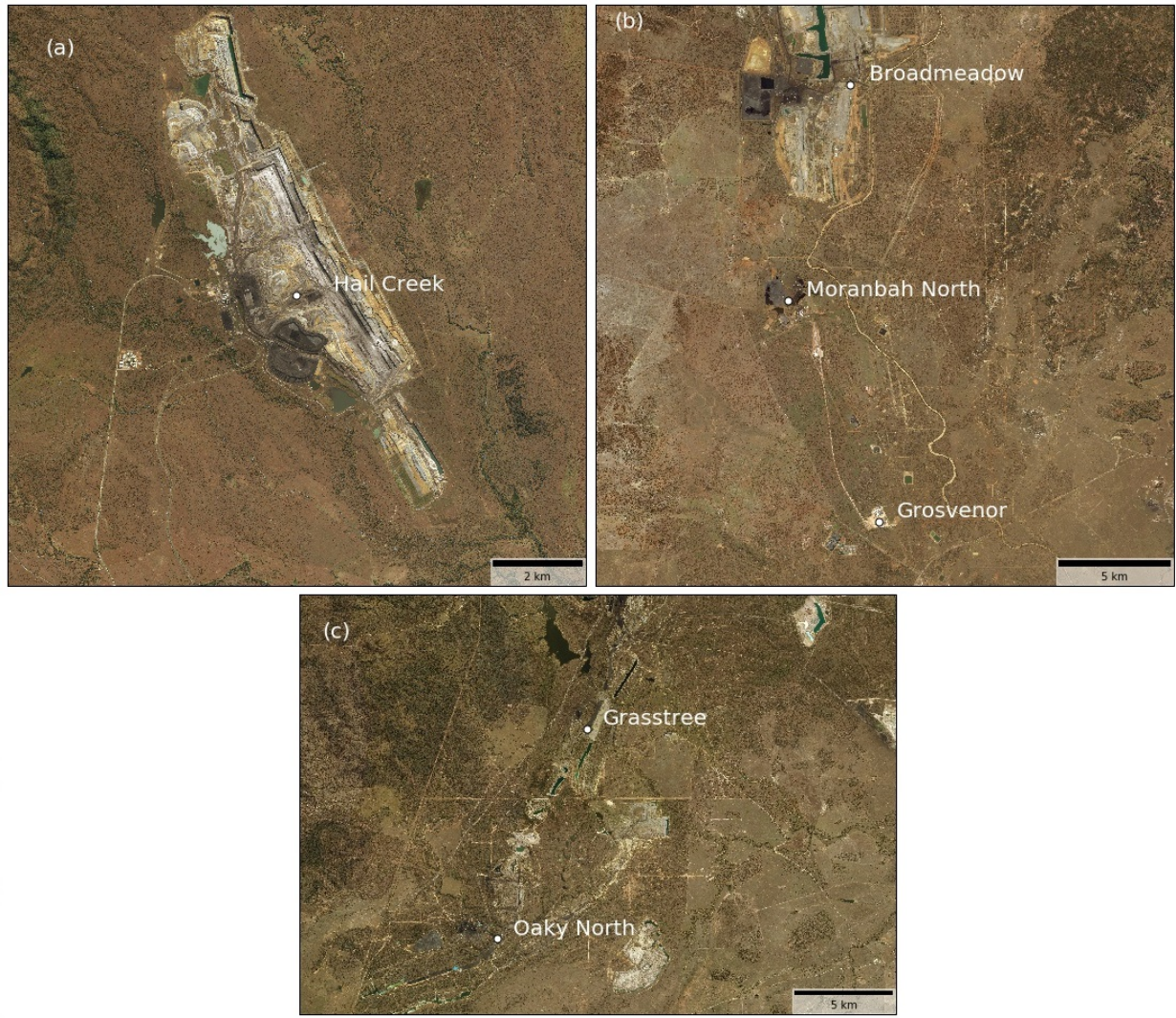

**Figure S4. Satellite imagery of coal mines at source 1, 2 and 3.**

Satellite images of (a) surface mine – Hail Creek, Source 1 (b) underground mines – Broadmeadow, Moranbah North and Grosvenor, Source 2 (c) underground mines – Grasstree and Oaky Creek, Source 3. Satellite imagery source: OpenStreetMap.

Location source from the below webpage, last accessed on 21st March 2021. (https://dsdip.maps.arcgis.com/apps/webappviewer/index.html?id=77b3197acf5f415cb6f24553d d16b9dc).

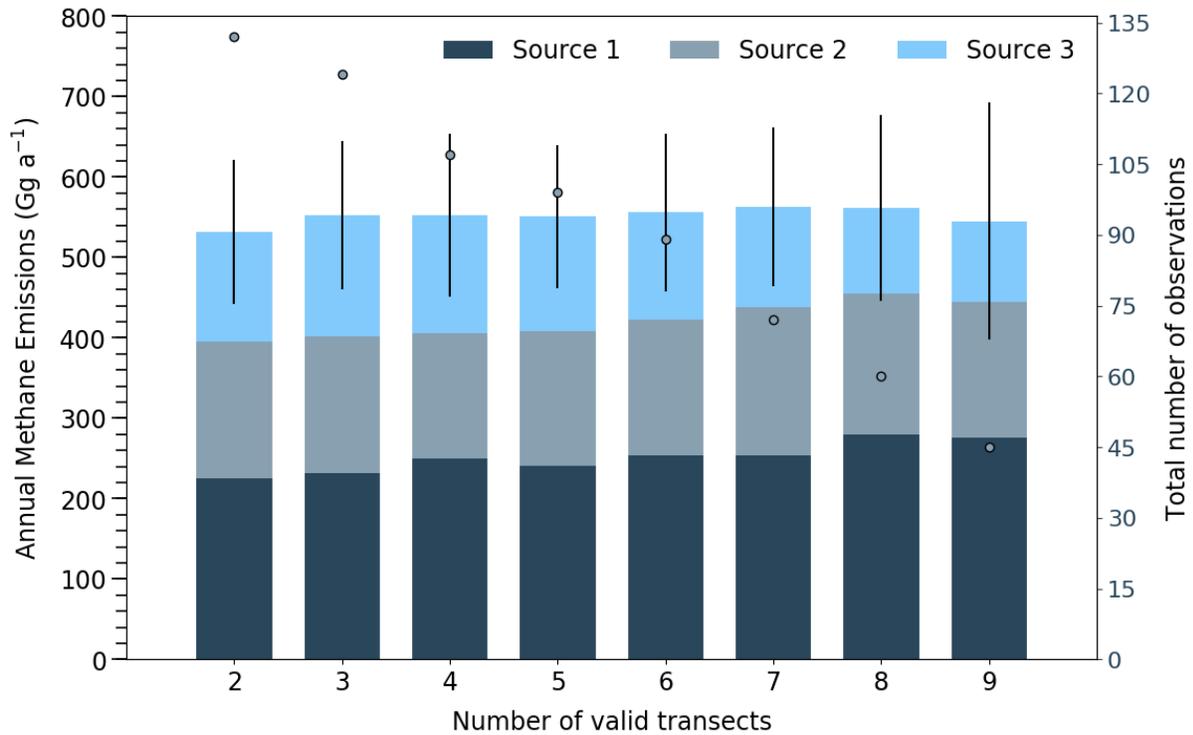

**Figure S5. Emissions sensitivity to number of valid transects.**

Total annual methane emissions quantified with varying number of valid transects are shown on left hand Y-axis and respective number of observations that were considered for the quantification are shown as scatter plot on right hand Y-axis. SRON Scientific TROPOMI XCH$_4$ data product with clear-sky observations ($q_a = 1$) and enhancement calculated using background upwind of the sources were selection criteria for this estimate. Our final estimate is provided using number of valid transects >= 3 resulting in 124 observations for the three sources. The uncertainty on the total is shown by 2σ (95% confidence interval).

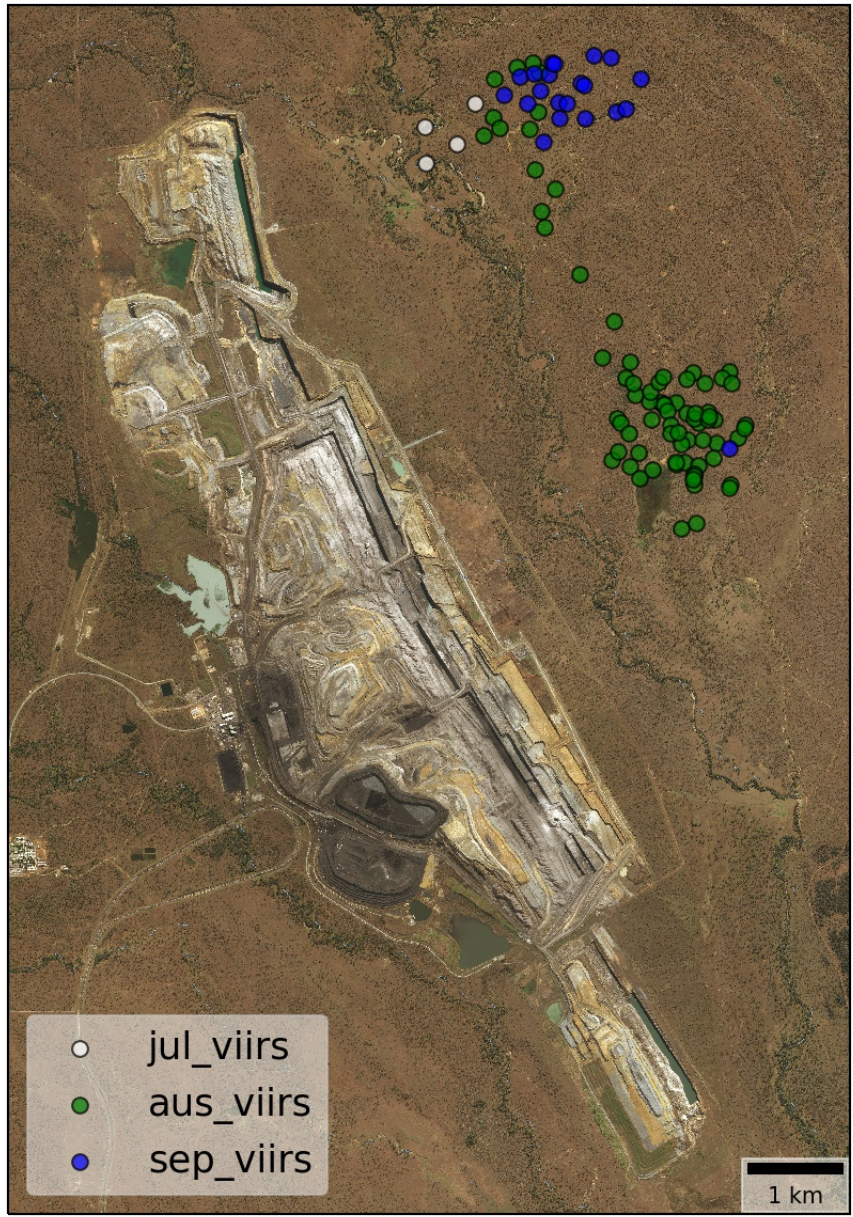

**Figure S6. Expansion activities over Hail Creek mine.**

High radiative power, flaring signals observed during the months of July, August and September of 2019 over the northeast area of the surface mine where the extension was approved[38]. Satellite imagery source: OpenStreetMap.

**Table S1:** Total methane emission estimates as sensitivity test to varying parameters.

| TROPOMI data | Quality flag and other varied conditions | Total number of valid plume observations for three source locations | Total Methane flux estimate (Gg a-1) |
|---|---|---|---|
| Scientific product[20] | $q_a = 1$, enhancement estimated using upwind background region | 124 | 570 ± 98* |
| | Data filters: aerosol optical thickness < 0.10, precision error < 10 ppb | 220 | 565 ± 80 |
| | $q_a = 1$, enhancement estimated using median $XCH_4$ of the study domain (20°S -24°S; 146°E - 150°S) as background | 124 | 550 ± 100 |
| Operational product[45] | $q_a = 1$, enhancement estimated using upwind background region | 144 | 525 ± 90 |
| | Data filters: aerosol optical thickness < 0.10, precision error < 10 ppb, | 258 | 500 ± 75 |

* baseline emission estimate.

**Table S2:** Annual coal production and fugitive methane emissions for coal producing states in Australia for 2018.

| Coal producing states | Coal production (ktonnne) | | Methane emissions (Gg) | |
|---|---|---|---|---|
| | Underground | Surface | Underground | Surface |
| Queensland | 48240[a] | 269119[a] | 343.88[e] | 196.26[e] |
| New South Wales | 64288[b] | 183832[b] | 366.53[e] | 61.06[e] |
| Victoria | | 45904[d] | | 0.50[f] |
| Tasmania | | 382[d] | | 0.26[f] |
| Western Australia | | 6649[d] | | 4.50[f] |
| **National** | **112528[c]** | **505886[c]** | **710.41[e]** | **261.58[g]** |

[a] Department of Natural Resources and Mines, Queensland, Summary of the raw and saleable coal of individual mines by financial year, https://www.data.qld.gov.au/dataset/27fefb68-dc98-4300-85b6-465f0df233a8/resource/9c3c1aaf-0afa-4e58-b67c-75c0d3574abd/download/production-by-individual-mines.xlsx.

[b] Raw and saleable coal production data compiled from annual review report of the respective individual coal mines for the year 2018.

[c] National raw coal production for year 2018, common reporting format (CRF) as reported to UNFCCC for year 2018.

[d] Raw coal production estimated for states of Victoria, Western Australia and Tasmania, by proportionately distributing the raw coal (calculated as difference between National and, Queensland and New South Wales) as per the saleable coal for the respective three states.

[e] Methane emissions reported to UNFCCC for 2018 (source: https://ageis.climatechange.gov.au/)

[f] Estimated using bottom-up emission factors available for surface mines in the national inventory report 2018, 2020 (Table 3.32) for Victoria and Tasmania state. For Western Australia, the emission factor similar to Tasmania was used for estimating emissions.

[g] Sum of surface based coal mine emissions.

**Table S3:** Australia's national greenhouse gas - methane emissions reported to United Nations Framework Convention on Climate Change (UNFCCC) for 2018.

| Sectors and sub-sectors | Methane emissions (Gg) |
|---|---|
| **A. Energy (combustion and fugitive)** | **1562.20** |
| Combustion | 77.92 |
| Fugitive emissions Solid fuels | 972.29 |
| Fugitive emissions Oil and natural gas | 511.98 |
| **B. Industrial processes and product use** | **3.10** |
| Chemical Industry | 0.58 |
| Metal Industry | 2.52 |
| **C. Agriculture** | **2335.52** |
| Enteric fermentation | 2066.73 |
| Manure management | 250.01 |
| Rice cultivation | 10.16 |
| Field burning of agricultural residues | 8.62 |
| **D. Land use, land use change and forestry incl. natural disturbances** | **630.73** |
| Forest land | 240.09 |
| Cropland | 0.71 |
| Grassland | 194.46 |
| Wetlands & Others[a] | 194.70 |
| Settlements | 0.77 |
| **E. Waste** | **480.48** |
| Solid waste disposal | 361.80 |
| Biological treatment of solid waste | 4.46 |
| Wastewater treatment and discharge | 114.22 |
| **Total CH$_4$ Emissions** | **5012.01** |

[a]Others include $CH_4$ from artificial water bodies.

All emissions are tabulated using common reporting format (CRF) table available at https://unfccc.int/documents/228034, last accessed on 21st March 2021.

**Table S4:** Implied emission factors estimated as ratio of emissions to coal production compared with studies at various domain levels.

| | Emission year | Implied emission factors (g/kg) | |
|---|---|---|---|
| | | **Underground** | **Surface** |
| **Australia (National)** | | | |
| EDGARv4.3.2[a,b] | 2012 | 10.23 | 0.55 |
| NIR 2017 (2019)[c] | 2012 | 8.38 | 0.45 |
| NIR 2018 (2020)[c] | 2012 | 8.38 | 0.45 |
| NIR 2017 (2019)[c] | 2017 | 6.05 | 0.47 |
| NIR 2018 (2020)[c] | 2017 | 6.05 | 0.47 |
| NIR 2018 (2020)[c] | 2018 | 6.31 | 0.52 |
| **Queensland (State)** | | | |
| NIR (2020)[c] | 2018 | 7.13 | 0.73 |
| **Reconstructed bottom-up, Sadavarte et al., 2021** | | | |
| Source 1 | | | 0.73 |
| Source 2 | 2018 | 8.60 | |
| Source 3 | | 5.73 | |
| **IPCC default (Global)[d]** | | | |
| < 200m | | 7.38 | 0.20 |
| 200-400m | | 13.87 | 0.88 |
| > 400m | | 19.62 | 1.49 |
| **Kholod et al., 2020 (Global)[e]** | | | |
| < 200m | | 10.08 | 2.03 - 3.38 |
| 200-400m | | 12.79 | |
| > 400m | | 14.62 | |
| brown coal | | | 0.52 |
| **TROPOMI (Individual source)** | | | |
| Source 1 | | | 34.12 |
| *Hail Creek* | | | |
| Source 2 | | | |
| *Broadmeadow* | 2018-2019 | 10.00 | |
| *Moranbah North: 270-370m[f]* | | | |
| *Grosvenor: 220m[g]* | | | |
| Source 3 | | | |
| *Grasstree: 370m[f]* | | 11.50 | |
| *Oaky North: 215-300m[h]* | | | |

**Supporting Animation 1.**
Four years (2017-2020) of Sentinel 2 satellite imagery over Hail Creek coal mine that shows the change in topography over the proposed coal mine extension area.